\documentstyle[draft]{mn}

\input psfig.sty

\begin{document}

\newcommand{\vecg}{\bmath{g}}
\newcommand{\vecG}{\bmath{G}}
\newcommand{\rhat}{\bmath{\hat r}}
\newcommand{\nhat}{\bmath{\hat n}}
\newcommand{\COBE}{{\sl COBE\/}}
\newcommand{\MAP}{{\sl MAP\/}}
\newcommand{\Planck}{{\sl Planck\/}}

\journal{Preprint UBC-COS-99-03, astro-ph/9906044}

%
\def\arreq{\kern-0.1truein &=& \kern-0.1truein}

\title[A Preferred-Direction Statistic for Sky Maps]
{A Preferred-Direction Statistic for Sky Maps}

\author[Emory F.~Bunn \& Douglas Scott]
{
	Emory F.~Bunn$^{1}$\thanks{ebunn@stcloudstate.edu}
	and Douglas Scott$^{2}$\thanks{dscott@astro.ubc.ca}\\
	\vspace*{1mm}\\
	$^1$Physics and Astronomy Department, MS 312,
720 4th Ave. S., St. Cloud State University, St. Cloud,
MN  56301-4498,~~U.S.A.\\
	$^2$Department of Physics \& Astronomy,
        University of British Columbia,
        Vancouver, B.C.~V6T 1Z1,~~Canada\\
}

\date{Accepted ... ;
      Received ... ;
      in original form ...}
\pubyear{1999}
\pagerange{000--000}
\maketitle

\begin{abstract}
Large patterns could exist on the microwave sky as a result of various
non-standard possibilities for the large-scale Universe --
rotation or shear, non-trivial topology, and single topological defects
are specific examples.  All-sky (or nearly all-sky)
CMB data sets allow us, uniquely, to constrain
such exotica, and it is therefore worthwhile to explore a wide range
of statistical tests.  We describe one such statistic here, which is based
on determining gradients and is useful for assessing the level of
`preferred directionality' or `stripiness' in the map.
This method is more general than other techniques for picking out specific
patterns on the sky, and it also has the advantage of being
easily calculable for the mega-pixel maps which will soon be available.
For the purposes of illustration, we apply this statistic to the four-year
\COBE\ DMR data.  For future CMB maps we expect this to be a useful
statistical test of the large scale structure of the Universe.  In principle,
the same statistic could also be applied to sky maps at other wavelengths,
to CMB polarisation maps, and to catalogues of discrete objects.  It
may also be useful as a means of checking for residual directionality
(e.g., from Galactic or ecliptic signals) in maps.
\end{abstract}


\section{Introduction}
The desire to understand the nature of the physical universe has driven
scientific inquiry throughout history.  An important aspect of this has been
the search for models which accurately describe the structure of the
Universe on the very largest scales.  Only in the last decade have we had the
means to access information from the largest of the scales that are
observationally within reach, namely those scales corresponding to the
light-crossing time in the age of the Universe.

In the absence of a unique mathematical description for the structure of
space, the only way to pin down its large-scale behaviour is through direct
observation of features which probe the largest scales.
Anisotropies in the Cosmic Microwave Background (CMB) give
us just such a diagnostic tool.  Furthermore,
since the detection of large-angle
anisotropies by the \COBE\ DMR experiment \cite{Smoot} it is now clear that
this is a tool that can be used in practice.  Although the \COBE\ detection
was impressive, the limited angular resolution and sensitivity make the
data set far from definitive.  Future CMB experiments, and in particular
the Microwave Anisotropy Probe (\MAP) and {\sl Planck} satellites will
provide data that will be far more constraining for the structure of the
Universe.

The traditional view is that most of the information content in the CMB
sky is contained in the angular power spectrum of anisotropies. As a
result, a great
deal of effort has gone into obtaining the best estimate of the power
spectrum from the \COBE\ DMR data \cite{COBE,Gor,WhiBun,Bond},
as well as for
other experiments.  However, it is certainly possible that the CMB sky
exhibits some large-scale pattern, not discernible from the power spectrum,
which would indicate some non-trivial large-scale structure.  Since
the endeavour of examining the Universe on its largest accessible scales is
such a grand one, it is certainly worthwhile to explore all statistics
which might be used to constrain such patterns.

Examples of {\it specific\/} such patterns
have already been investigated, although we
would like to stress that the current list is likely to be far from exhaustive,
and the best approach is the empirical one: simply look to see if there are
surprises in the data.  Already much effort has gone into examining the
consequences of non-trivial topology, with a repetition scale close to the
Hubble radius today
\cite{SteScoSil,deOSmoSta,LevBBS,CorSpeSta,SouPogBon}.  One manifestation of
a short periodicity scale in two direction is that the sky would appear to be
composed of concentric `stripes,' superimposed on the background of
fluctuations \cite{Sta}.  Another class of possibilities are universes
with rotation or shear, which define a preferred direction on the sky around
which there can be special patterns \cite{BarJusSon};
specific Bianchi models have been investigated in detail
\cite{BunFerSil,KogHinBan}.
A further example is the possibility that single large-scale topological
defects could give rise to particular patterns, e.g., a step-like
discontinuity caused by the largest cosmic string \cite{Per}, or a localised
pattern caused by the
last texture to unwind \cite{Mag}.  It is also possible that properties
of particle fields in the Universe could create electromagnetic effects
yielding special directions (see e.g.~Carroll \& Field~1997 for a recent
discussion).
These examples represent the cases already studied, where there are known
patterns which exhibit some directionality.  We wish to consider the idea
of searching for general phenomena which might give rise to a preferred
direction on the sky, with no particular prejudice about the nature of
the pattern.

The closely related general topic of non-Gaussianity, i.e.~the correlation
between phases of the modes, has also received a great deal of recent
attention
\cite{FerMagGor,WanHivGor,LewAlbMag,PanValFan,HobJonLas,NovFelSha,Hea}.
Many of these studies have made
use of statistics calculated in the Fourier domain.  However,
there are significant advantages in developing statistics
which can be calculated directly from pixel values, and which are designed
to search for specific forms of non-Gaussianity.  Of course non-Gaussianity
could take many forms, each requiring a differently tailored statistical
test.  We consider our specific statistic to be
part of the arsenal of statistics for assessing the large-scale
structure of space-time which might involve a preferred direction
on the CMB sky.  Indeed we would claim it is the obvious statistic to
use for this purpose, as we describe in the next section.

\section{A preferred-direction statistic}

We want to devise a statistic that will tell us whether a sky
map like that obtained from \COBE\
has a preferred direction.  In particular, we would like to determine
whether temperature contours on an all-sky map tend to be
either elongated or squashed along some particular direction
in space.

To pick out such a direction, it seems most natural to work with the
gradient $\bmath{\nabla}T$ of the temperature map.  In a noisy map, of
course, some amount of smoothing will be necessary before $\bmath{\nabla} T$
is estimated.  In particular, suppose that we have a sky map with
$N_{\rm p}$ pixels.  We will estimate the gradient of the map by grouping
the pixels into $N_{\rm b}$ localised blocks and fitting a linear function
to the pixels in each block.  In this manner, we obtain a gradient vector
$\vecg_p$ for $p=1,\ldots,N_{\rm b}$; i.e., we find an estimate of the
smoothed gradient of $T$ corresponding to each of the $N_{\rm b}$ blocks of
pixels.  By varying the number of blocks $N_{\rm b}$ (and hence the size of
each block), we can control the amount of smoothing.  The
appropriate amount of smoothing depends on the resolution
and the noise level of our data set and must be determined
by trial and error on mock data sets.

The precise method for dividing the $N_{\rm p}$ pixels into $N_{\rm b}$
blocks will depend on the exact pixelisation scheme.
As we will see below, the \COBE\ sky cube pixelisation leads
to a natural method for grouping pixels: we simply subdivide
each face of the cube into squares.

Let us suppose we have estimated $\bmath{\nabla} T$ at each of $N_{\rm b}$
locations.  Let $\vecg_p$ be our estimate of the gradient at location
$p$.  Naturally, $\vecg_p$ must be tangent to the surface of the
celestial sphere at that location.  If $\rhat_p$ is a unit vector
pointing toward the location of gradient estimate $p$ (specifically,
say the centroid of the pixels that were used to make
that estimate), then $\vecg_p$ must
be perpendicular to $\rhat_p$, i.e., 
\begin{equation}
\vecg_p\cdot\rhat_p=0.
\label{eq:perpen}
\end{equation}
We assume that we have estimated the gradient at $N_{\rm b}$ locations,
so that our data set consists of the $N_{\rm b}$ vectors
$\vecg_1,\ldots,\vecg_{N_{\rm b}}$.  If there
is a preferred direction in the map, then $\vecg$ should
tend either to line up with some particular direction or to avoid
some particular direction.  To see whether or not this happens,
we should compute
\begin{equation}
f(\nhat)=\sum_{p=1}^{N_{\rm b}} w_p(\vecg_p\cdot\nhat)^2
\label{eq:fdef}
\end{equation}
for all unit vectors $\nhat$.  The weights $w_p$ should be chosen
to avoid introducing
spurious signals
into the statistic; we will describe
how to do this below.
If our map has a preferred direction $\nhat$, then $f(\nhat)$ should be either
anomalously small or anomalously large compared to other directions.
That is, a large value of the `directionality' ratio
\begin{equation}
{\cal D}\equiv {\max_{\nhat} f(\nhat)\over\min_{\nhat} f(\nhat)}
\end{equation}
will alert us to the presence of a preferred direction.  

Of course, the ratio of maximum to minimum values is not the only
statistic one could derive from the function $f$, although, since
$f$ is a very simple function (a quadratic in $\nhat$), 
the number of essentially distinct statistics that can be derived from
it is limited.  We choose this particular statistic because it is
simple, and because it picks out maps for which $f$ has either
extremely high peaks or extremely deep valleys.  A preferred direction
in a sky map could manifest itself either through very large or very
small values of $f$: for instance, a model that is topologically small
in one dimension will produce a very low value for $f$ when $\nhat$
points along the preferred direction, while one that is small in two
dimensions will produce a high value.  Both sorts of maps
are detected by the statistic $\cal D$.

A major advantage of this statistic over
other methods of finding preferred directions
is that it is extremely easy to compute.
$f$ is quadratic in $\nhat$, which means that finding its extrema
for a given sky map involves nothing more than solving a 3-dimensional
eigenvalue problem.  We will work this out in detail below, but
first let us consider how to choose the weights $w_p$.

If the pixels were uniformly distributed over the sphere, of course
we would want uniform weights.  But invariably we have non-uniform sky
coverage, requiring non-uniform weights.  To see why uniform
weights are unacceptable, consider the
standard \COBE\ situation where a region near the equator has been cut
out.  That means that the pixel directions $\rhat_p$ will have $z$
components whose absolute values are larger,
on average, than their $x$ and $y$ components (these components
are of course specified with reference to an ordinary
Cartesian coordinate system
with the $z$-axis perpendicular to the equatorial plane).
According to
equation (\ref{eq:perpen}), that means that $\vecg$ will have
$z$ components that are closer to zero than they
would be if we had full sky coverage.  So $f(\bmath{\hat z})$
will be smaller than $f(\bmath{\hat x})$ or $f(\bmath{\hat y})$.
That is dangerous,
because it means that the $z$ direction will look like a preferred
direction even when it is not.  The solution is to use non-uniform
pixel weights;  in this particular case, we would give the pixels near the
equator higher weights than those near the poles.

We must now work out a prescription for assigning pixel weights
that will avoid `false positive' results of the sort described above.
Consider the null hypothesis that there
is in fact no preferred direction.  Then each $\vecg_p$ is a random
vector with an isotropic distribution in the tangent plane to the
sphere at that point.  That is, $\vecg_p$ must be in the plane
perpendicular to $\rhat_p$, but within that plane it is equally likely
to point in any direction (for standard Gaussian theories, $\vecg_p$
will be a two-dimensional Gaussian random vector, but for the moment we do not
need to assume that).  We want to choose the weights $w_p$ so that
the function $f$ has no particular preferred direction in this case.
Specifically, we require that the ensemble-average quantity $\langle
f\rangle$ be constant as a function of $\nhat$.

We can write equation (\ref{eq:fdef}) as
\begin{equation}
f(\nhat)=\nhat^T{\bf A}\,\nhat,
\label{eq:fmat}
\end{equation}
where the $3\times 3$ matrix ${\bf A}$ has elements
\begin{equation}
A_{ij}=\sum_{p=1}^{N_{\rm b}} w_p\,g_{pi}\,g_{pj},
\end{equation}
and $g_{pi}$ is the $i$th Cartesian coordinate of $\vecg_p$.  Then
$\langle f(\nhat)\rangle$ will be independent of $\nhat$ if and
only if $\langle {\bf A}\rangle$ is proportional to the identity matrix.
We have not decided how to normalise $f$ yet, so we may as well use
that freedom to demand that it be {\it equal} to the identity:
\begin{equation}
\langle A_{ij}\rangle=\delta_{ij}.
\label{eq:aconstraint}
\end{equation}

Since 
\begin{equation}
\langle A_{ij}\rangle=\sum_pw_p\langle g_{pi}g_{pj}\rangle,
\label{eq:avgea}
\end{equation}
equation
(\ref{eq:aconstraint}) constrains the possible choices of weights $w_p$.
To cast this constraint in a useful form, we have to make use of the
assumed isotropy of $\vecg_p$.  Let $\vecG_p$ be a 3-dimensional
isotropic random vector, and choose $\vecg_p$ to be its projection
on to the tangent plane:
\begin{equation}
\vecg_p=\vecG_p-(\vecG_p\cdot\rhat_p)\rhat_p
\label{eq:gproj}
\end{equation}
-- this is just a simple and convenient way to impose the requirement that
$\vecg_p$ be isotropic in the tangent plane.  Since $\vecG_p$ is isotropic,
we must have $\langle \vecG_p\rangle=0$ and $\langle G_{pi}G_{pj}\rangle
=P_p\delta_{ij}$, where $P_p$ is the mean-square amplitude
of $\vecG_p$.

Then application of equation (\ref{eq:gproj}) yields
\begin{eqnarray}
\langle g_{pi}g_{pj}\rangle\arreq
\langle G_{pi}G_{pj}\rangle-r_{pi}\sum_{\alpha=1}^3\langle G_{p\alpha}
G_{pj}\rangle r_{p\alpha}\nonumber\\
& &\qquad -r_{pj}\sum_{\beta=1}^3\langle G_{pi}G_{p\beta}
\rangle r_{p\beta} \nonumber\\
& & \qquad
+\left(\sum_{\alpha,\beta=1}^3\langle G_{p\alpha}G_{p\beta}
\rangle r_{p\alpha}r_{p\beta}\right)r_{pi}r_{pj} \\
\arreq
P_p(\delta_{ij}-r_{pi}r_{pj}).\nonumber
\end{eqnarray}

Combining this with equations (\ref{eq:avgea}) and (\ref{eq:aconstraint}),
we find that
\begin{equation}
\delta_{ij}=\sum_{p=1}^{N_{\rm b}} w_pP_pQ_{pij},
\label{eq:constraints}
\end{equation}
with
\begin{equation}
Q_{pij}=\delta_{ij}-r_{pi}r_{pj}.
\end{equation}
Since
equation (\ref{eq:constraints}) is symmetric, it
gives six constraints on the $N_{\rm b}$ pixel
weights $w_p$.  Of course, $N_{\rm b}\gg 6$, so
the choice of weights is still heavily
underdetermined, and we need an additional criterion to fix them.
We choose to adopt
the simple and reasonable criterion that the weights should
be as nearly equal as possible, subject to (\ref{eq:constraints}).
That suggests minimizing the variance of $\{w_p\}$.  In fact, it is
somewhat more convenient to minimize the
variance of $\widetilde w_p\equiv w_pP_p$, so
let us do that instead.\footnote{Why $\widetilde w_p$ instead of $w_p$?
There are three reasons: (1) as we shall see, the mathematics is slightly
simpler; (2) $P_p$ is proportional to the mean-square value of $\vecg_p$,
which contains contributions from both signal and noise -- typically,
any variation of $P_p$ from pixel to pixel will be due to noise variation,
so large values of $P_p$ will correspond to noisy patches, and it can't hurt
to down-weight noisy patches a little; (3) as we shall see below, we are
going to do quite a bit of smoothing before estimating the gradient,
so noise variation won't matter much, and $P_p$ will be essentially
constant from pixel to pixel anyway.}

We want to minimize
\begin{equation}
\mbox{Var}(\widetilde w)={1\over {N_{\rm b}}}\sum_{p=1}^{N_{\rm b}}
  \widetilde w_p^2-\left({1\over {N_{\rm b}}}\sum_{p=1}^{N_{\rm b}}
  \widetilde w_p\right)^2\!.
\end{equation}
Taking the trace of
equation (\ref{eq:constraints}), we find that $\sum_p\widetilde w_p={3\over 2}$,
so the second term in the above equation is constant.
(This is the reason it's simpler to work with $\widetilde w_p$ than with
$w_p$.)
Hence, all we have to do is minimize
\begin{equation}
\Delta^2\equiv {1\over 2}\sum_{p=1}^{N_{\rm b}}\widetilde w_p^2
\end{equation}
subject to the constraints (\ref{eq:constraints}).
Introducing a symmetric $3\times 3$ matrix $\Lambda$ of Lagrange
multipliers, this minimization problem becomes
\begin{equation}
\widetilde w_p=\sum_{i,j=1}^3 \Lambda_{ij}Q_{pij}.
\label{eq:solveforw}
\end{equation}
Plugging this back into the constraint equation (\ref{eq:constraints}),
we get
\begin{equation}
\delta_{ij}=\sum_{k,l=1}^3\Lambda_{kl}\widetilde Q_{ijkl},
\label{eq:solveforlambda}
\end{equation}
where
\begin{equation}
\widetilde Q_{ijkl}\equiv\sum_{p=1}^{N_{\rm b}} Q_{pij}Q_{pkl}.
\end{equation}
Equations (\ref{eq:solveforlambda}) form a 6-dimensional linear
system that is easily solved for $\Lambda$.  Once $\Lambda$ is known,
equation (\ref{eq:solveforw}) yields the weights $\widetilde w_p$.

Once we have chosen the weights, obtaining the extrema
of $f$, and hence the statistic ${\cal D}$, is extremely rapid.
Let $\nhat$ be a unit vector that points in the direction
of one of the extrema.  To find $\nhat$, we must
extremise $f$ subject to the constraint that $\nhat$ have magnitude
one.  This constrained extremisation problem is
solved by introducing a Lagrange multiplier:
for each of the three Cartesian
components $\hat n_i$ of $\nhat$, we set the derivative of $f$ with respect to 
$\hat n_i$ equal to 
the derivative of the constraint equation $\sum_{i=1}^3\hat n_i^2=1$,
multiplied by the Lagrange multiplier $\lambda$:
\begin{equation}
2\sum_{j=1}^3A_{ij}\hat n_j=\lambda (2\hat n_i).
\end{equation}
This set of three equations is equivalent to the matrix equation
\begin{equation}
{\bf A}\,\nhat=\lambda\,\nhat.
\end{equation}
In other words, the locations of the extrema of $f$ are eigenvectors
of ${\bf A}$.  Furthermore, the extreme values themselves are given by the
eigenvalues.

The $3\times 3$ symmetric matrix ${\bf A}$ always has three eigenvectors,
so $f$ has three critical points, one maximum, one minimum, and one saddle
point.  Once the elements of ${\bf A}$ have been computed (requiring
$O(N_{\rm p})$ operations), all we have to do to find ${\cal D}$ is 
solve a $3\times 3$ matrix eigenvalue problem and take the ratio
of the largest and smallest eigenvalues.

\section{Comparison with other methods}
We can now search the \COBE\ DMR data set for specific directionality
and compare our results with other approaches.  
We adopt as our basic data set a weighted average of the 53 and $90\,$GHz
A and B sky maps, in ecliptic pixelisation, with pixels
near the Galactic plane removed following the \COBE\ DMR
group's `custom cut' \cite{COBE}.  

A best-fitting monopole and dipole
are removed from the data set at the beginning of the analysis.
Dipole removal is absolutely crucial in nearly
all large-angle CMB analysis \cite{TegBun}, but it is more than
usually important in this case, since the statistic
$\cal D$ is extremely sensitive to the dipole.
(If we didn't remove the dipole, the function $f$ would have an enormous peak
oriented along the dipole axis.)
Fortunately, least-squares fitting removes all trace of the dipole:
two maps that differ only by a dipole will look identical after least-squares
dipole removal.  

\setcounter{figure}{0}
\begin{figure}
\begin{center}
\psfig{file=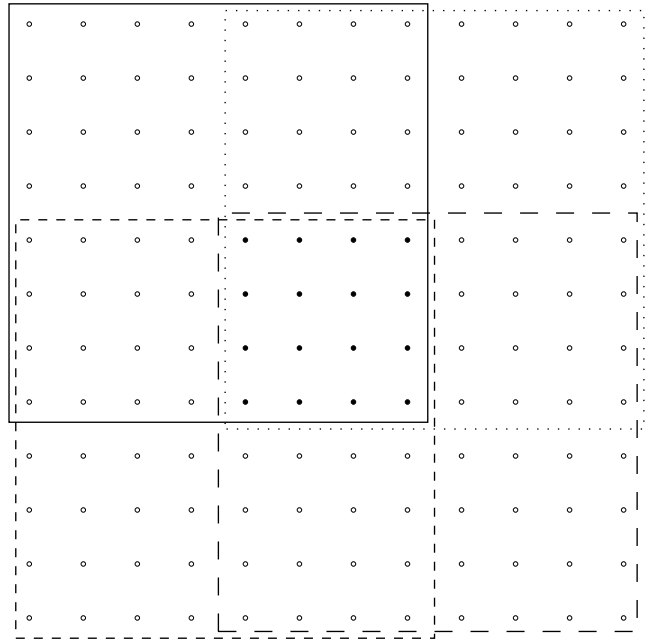,height=8.5truecm,angle=0}
\caption{How pixels are grouped into blocks.
The dots represent some of the pixels on one face of the \COBE\ sky cube.
The four boxes show the four overlapping blocks to which
the 16 central pixels 
(represented by the solid circles) belong.  In this manner,
all pixels belong to multiple blocks.
Any blocks that extend off the edges of the face continue
to adjacent faces.  In addition, there are eight blocks that cover
the corners of the sky cube -- each of these eight blocks 
has only 48 pixels, rather than 64, since it consists
of one $4\times 4$ `sub-block' on each of the three cube faces
that meet at that corner.}
\label{fig:blocks}
\end{center}
\end{figure}

In order to compute a smoothed gradient map, we must group pixels
together into localised `blocks,' with each block providing one
gradient estimate.  The pixelisation scheme for the DMR maps, in which
the sky is divided into six faces, with each face subdivided into a
$2^5\times 2^5$ square of pixels, suggests a natural way of grouping
pixels: we subdivide each face into square blocks with $2^n$ pixels on
a side.  The appropriate value of $n$ (i.e., the best smoothing
level) is found by Monte Carlo simulations: $n$ should be chosen to
maximise the probability with which maps that contain a preferred
direction can be distinguished from those that do not.\footnote{In order
to avoid biasing the final results, it
is essential that choices like this one be made solely on the basis
of simulated maps, without looking at the real data.}
The simulations, which will be described in more detail below, 
lead to the conclusion that maximum sensitivity is
typically achieved when $n=3$, meaning that each gradient estimate is
computed by fitting a linear function to a square block of 64 pixels.
The blocks need to be large because the \COBE\ signal-to-noise
ratio is relatively low.

With such large blocks, we have relatively few gradient
estimates to work with.  In order to increase the
number of data points, we allow the blocks to overlap,
so that each pixel is contained in four blocks as shown in
Fig.~\ref{fig:blocks}.\footnote{Naturally, this scheme induces
correlations between different gradient estimates -- specifically,
adjacent estimates are anticorrelated.  However, this is harmless,
since our statistic is calibrated with Monte Carlo simulations that
automatically account for these correlations.}
With this arrangement, there
are 378 blocks in a full-sky map, but only 199 remain after the
Galactic cut is applied (we choose to reject a block if the cut
removes more than 20 per cent of its pixels).

\begin{figure}
\begin{center}
\psfig{file=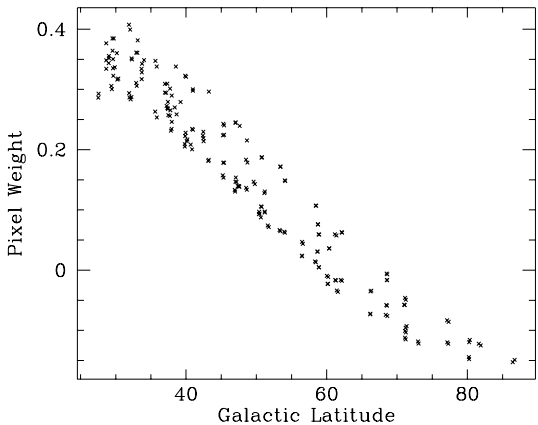,height=6truecm,angle=-90}
\caption{Pixel weights.
These are our calculated weights for blocks of pixels in the \COBE\ map,
using the `custom cut' of the Galactic plane.}
\label{fig:weights}
\end{center}
\end{figure}

Now that we know the pixel locations, we can compute the weights
$w_p$ to be used in the preferred-direction search.  The weights
are plotted in Fig.~\ref{fig:weights} as a function of Galactic
latitude.\footnote{In order to compute the weights, we need to know
$P_p$, the mean-square value in the $p$th gradient estimate.
These values depend weakly on the cosmological model chosen.
For Fig.~\ref{fig:weights}, the weights were computed using
a standard $n=1$ Harrison-Zel'dovich spectrum with no preferred
direction, but there is no significant qualitative change in the
plot for other reasonable models.}
As expected, the Galactic cut causes pixels near the poles to
be given less weight than those near the equator; in fact, about
23 per cent of the pixels have negative weight.

\begin{figure}
\begin{center}
\psfig{file=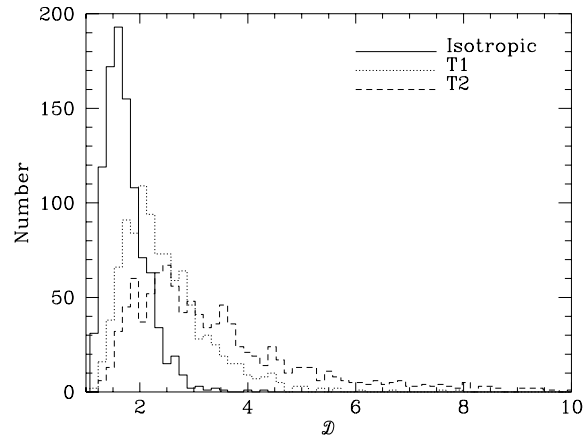,height=6truecm,angle=-90}
\psfig{file=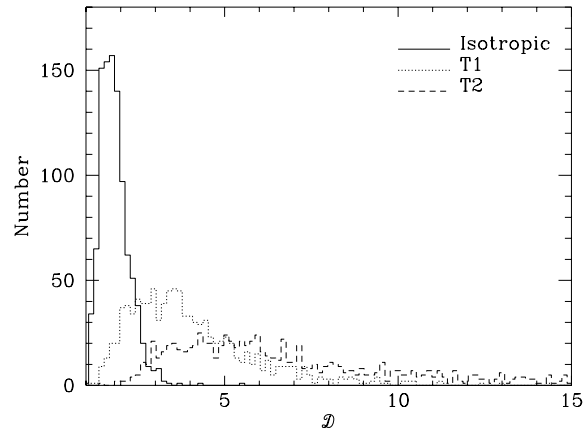,height=6truecm,angle=-90}
\caption{The distribution of the directionality statistic, ${\cal D}$.
These histograms show the distribution of the statistic ${\cal D}$
in Monte Carlo simulations.  The three cases shown 
are an isotropic Gaussian model and toroidal
models with one and two `small' dimensions of
0.3 times the distance to the last-scattering surface (labelled $T_1$
and $T_2$, respectively).
The top panel shows results of simulations with \COBE\
signal-to-noise levels; the signal-to-noise in the
bottom panel was inflated by a factor of five.}
\label{fig:hists}
\end{center}
\end{figure}

In order to assess the efficacy of the ${\cal D}$
statistic in distinguishing models with a preferred direction
from those without, we chose to focus on a specific model with preferred
direction -- we performed Monte Carlo simulations
of flat toroidal Universes.  When all three torus dimensions
are large, the temperature distribution is, to an excellent approximation,
an isotropic Gaussian random field, with no preferred direction.
When one or two dimensions are made small, though, a preferred direction
emerges.

The upper panel of
Fig.~\ref{fig:hists} shows the results of some of these
simulations.  One histogram shows the distribution of ${\cal D}$
for a standard, isotropic, Gaussian model with an $n=1$
Sachs-Wolfe power spectrum, $C_l\propto 1/l(l+1)$.  The other
two histograms represent models with a preferred direction.
The plot labelled `T1' is a model that is topologically
small in one dimension and large in the other two.  The
small dimension has a length of 0.3 times the comoving distance 
to the last-scattering surface.  The plot labelled `T2' is
as small as this in two dimensions.  All of these simulations were made with
signal-to-noise ratios that match that of the DMR data.

As expected, the topologically small models do have higher
values of ${\cal D}$ on average.  Unfortunately, the
distribution is quite broad, reducing the statistic's ability
to distinguish the isotropic model from the anisotropic ones (although
for much shorter small dimensions the directionality would be easily noticed).
This is due in large part to the poor signal-to-noise ratio
in the DMR data, as the lower panel of Fig.~\ref{fig:hists}
shows.  This figure was produced in the same way as the upper
panel, except that the signal-to-noise ratio was increased
by a factor of five.

\begin{table}
\begin{center}
\caption{Probability (in per cent) 
of detecting a topologically small Universe at
95 per cent confidence in Monte Carlo simulations.
The models have one or two small directions (T1 and T2,
respectively) with \COBE\ noise, and using either the \COBE\ signal
normalisation, or 5 times that value.
\label{table}
}
\begin{tabular}{|l|l|l|l|l|}
\noalign{\hrule}
Length & T1 & T2 & T1 & T2\\
scale & \COBE\ & \COBE\ & $5\times$ \COBE\ & $5\times$ \COBE\ \\
      & norm.  & norm.  & norm. & norm.\\
\noalign{\smallskip}
0.3 &  51.2 & 75.6 &  74.9  & 95.8\\
0.5 &  22.4 & 35.5 &  32.8  & 43.7\\
0.8 &   8.0 &  7.2 &   6.0  &  9.2\\
\noalign{\hrule}
\end{tabular}
\end{center}
\end{table}

Table~\ref{table} shows one way of quantifying the power of this
statistic.  We created simulated sky maps for both an isotropic
Gaussian model and a set of toroidal models with one and two
small dimensions.  We computed the value of ${\cal D}$ for
all of the realisations.  For each anisotropic sky, we then
asked the following question: is the value of ${\cal D}$ produced
by this sky large enough to rule out the isotropic model
at 95 per cent confidence?  In other words, we determined whether
the given value of ${\cal D}$ was greater than the 95th percentile
of the isotropic values.  For each anisotropic model,
the fraction of the time that this
occurs is the probability that a preferred direction would
be detected (i.e., that isotropy would be ruled out)
at 95 per cent confidence.

Table~\ref{table} shows these detection probabilities
for both T1 and T2 models with length scales of 0.3, 0.5, and
0.8 times the distance to the last-scattering surface, and
for signal-to-noise levels equal to the \COBE\ value and five
times the \COBE\ value.  In order to see how well our statistic
compared with others, we implemented the `$S$' statistic described in
\cite{deOSmoSta} and performed similar simulations
with this statistic.  We find that the $S$ statistic
is more powerful than ours: for instance, when the $S$
statistic is used, the last row of table~\ref{table}
changes to 11.4, 20.6, 15.8, 25.0 per cent.
However, we should stress that this is precisely the case for which
$S$ was developed; our statistic is much more general,
and we will discuss its utility in the next section.

\begin{figure}
\begin{center}
\psfig{file=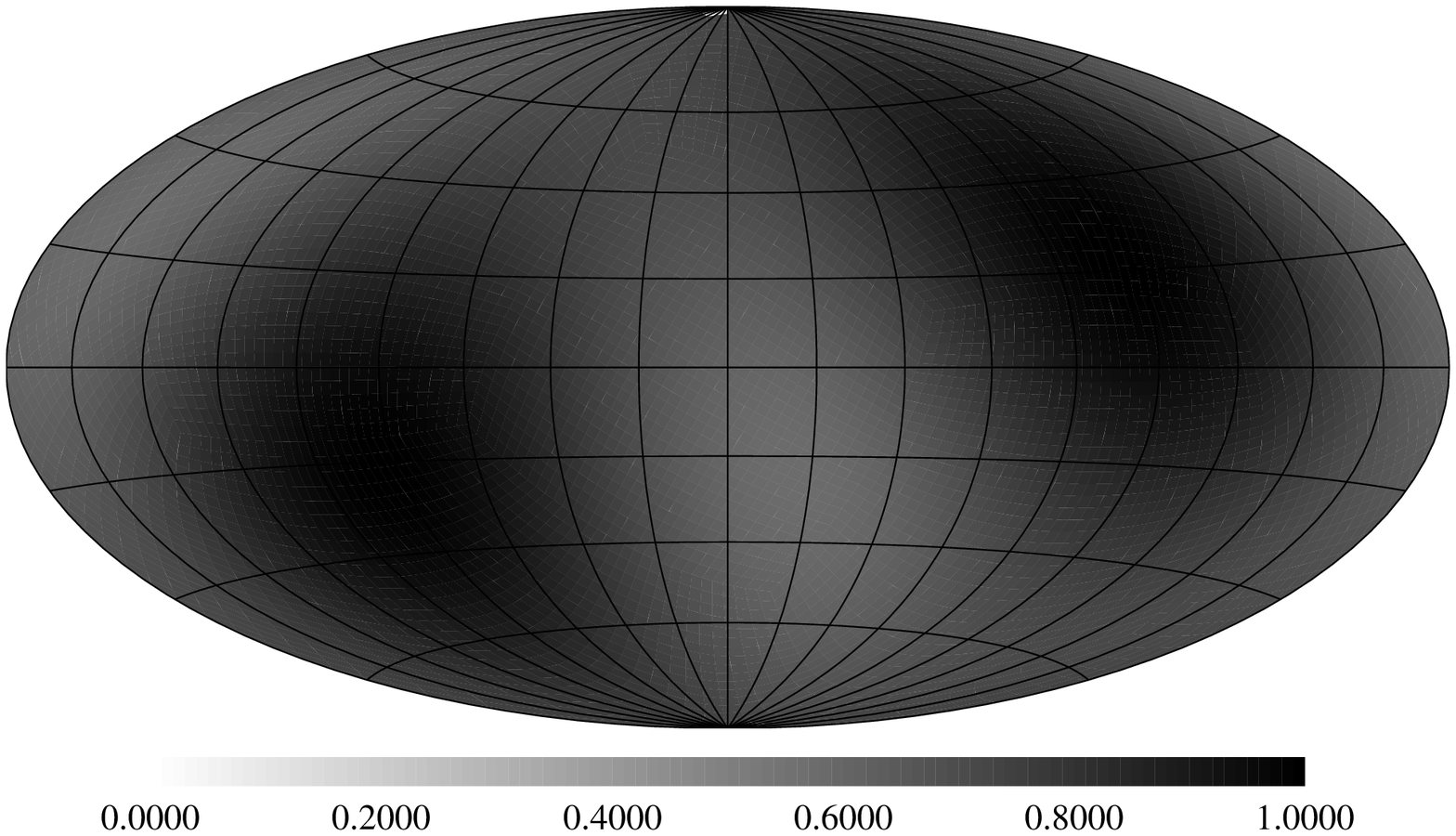,height=5truecm,angle=0}
\psfig{file=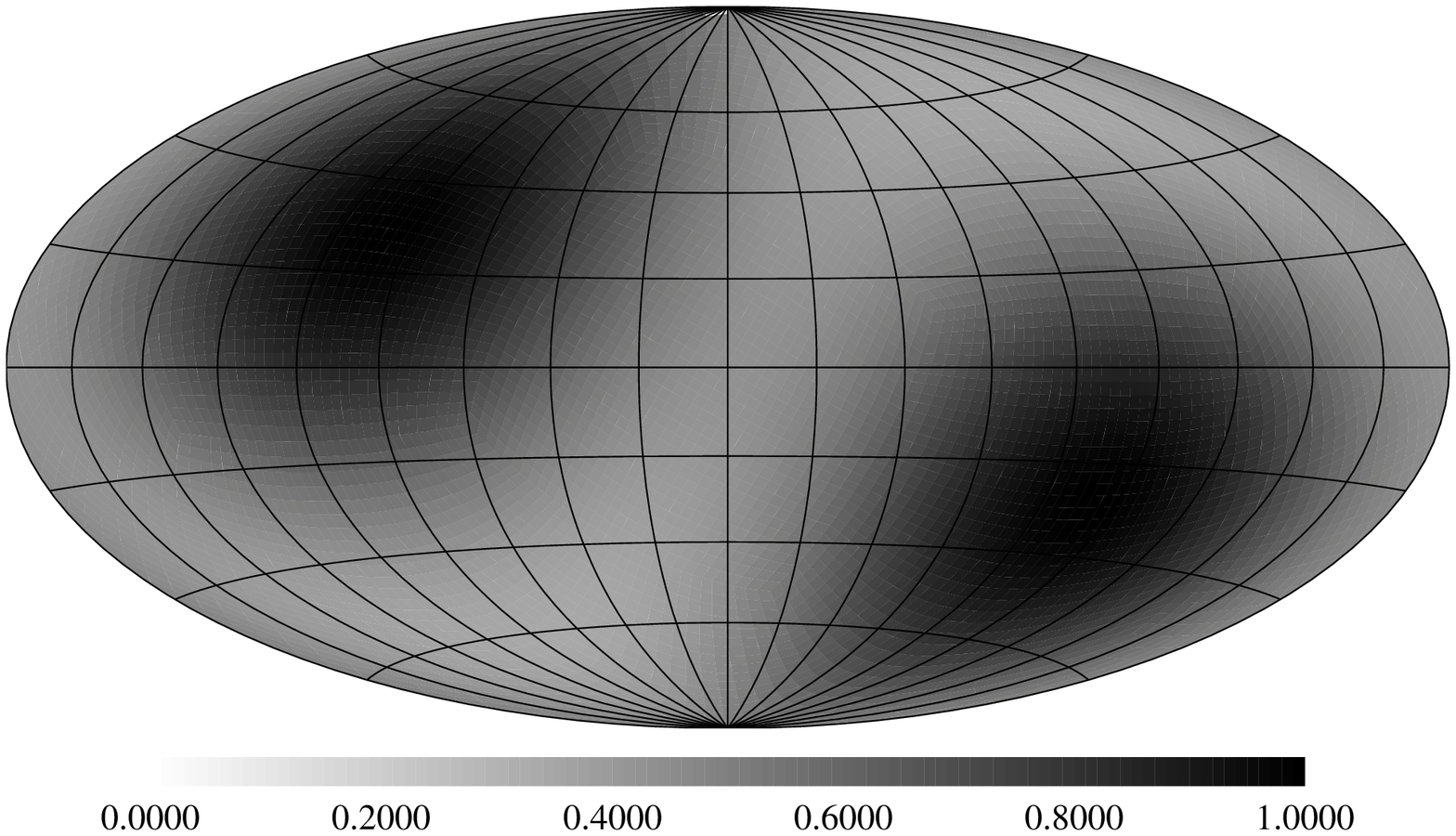,height=5truecm,angle=0}
\caption{Maps of $f/f_{\rm max}$ in Aitoff-projected Galactic coordinates.
The Galactic centre is at the centre of the maps; Galactic longitude
increases from right to left.
The real DMR data were used to produce the top panel;
for the bottom panel, a large prominent hot spot, with an
amplitude of 200$\,\mu$K and a $20^\circ$ radius, centred at
$(l,b)=(45^\circ,45^\circ)$,
was added to 
the data.  The grey-scale runs from zero to one, so the low level
of contrast in the top panel illustrates that $\cal D$ is
not very large ({\it i.e.}, that $f_{\rm min}/ f_{\rm max}\equiv
1/{\cal D}$ is not very small).}
\label{fig:fmaps}
\end{center}
\end{figure}

We have focused primarily on the results of simulations, since
our main interest is in assessing the power of the statistic.
However, for completeness we now mention the limits
that can be placed on topologically small Universes by
applying this statistic to the {\it actual} DMR data.
The value of $\cal D$ for the real data is 1.74.
The maximum value of $f$ occurs at
Galactic coordinates
$(l,b)=(84^\circ,-22^\circ)$, and the minimum is at
$(343^\circ,-26^\circ)$.\footnote{Of course,
in both cases the antipodal point could just as 
easily have been chosen: $f(\nhat)=f(-\nhat)$.}  
A map of $f$ for the real data is shown in Fig.~\ref{fig:fmaps}.
For comparison,
a map of $f$ for a sky with a prominent hot spot added is also shown.
(The dullness of these maps is not surprising: since $f$ is
a quadratic form, it cannot have much structure.)

The fact that the directions found for the \COBE\ data
are not particularly close to various
`special' directions in the sky is reassuring.  (Even more reassuring
is that the maxima and minima of $f$ are distributed isotropically
over the sky in our simulations.)  The maximum value of $f$ occurs
$68^\circ$ from the Galactic pole, $84^\circ$ from the Galactic
centre, and $26^\circ$ from the CMB dipole \cite{Dipole}.  
The minimum value occurs
at $64^\circ$, $31^\circ$, and $78^\circ$ from these three directions.

The directions picked out by our statistic do not lie especially
close to prominent hot or cold spots in the data.  The three
most significant spots in the DMR data, as listed by Cay\'on \& Smoot (1995),
are all at least $35^\circ$ from the locations of both the maximum and minimum
values of $f$.  

Bromley \& Tegmark (1999) have
identified another potentially significant direction in the \COBE\
data.  They point out that a recent claimed detection of non-Gaussianity
in \COBE\ \cite{FerMagGor} can be made to disappear if
pixels near $(l,b)=(257^\circ,39.5^\circ)$ are excised from the data.
This direction is about $18^\circ$ from the maximum value of $f$.
While this is a closer match than any of the other directional comparisons
made above, it is not close enough to
be statistically significant, and we suspect it is a mere
coincidence.\footnote{In the past three paragraphs, we have
compared the location of the maximum and minimum of $f$ to seven
locations on the sky.  If these had been seven randomly-chosen
directions, the probability of getting a match to within
$18^\circ$ would have been about 50\%.}

We can use the value of $\cal D$ for the real data to rule out
various models in the usual way: if a model produces $\cal D$-values
larger than the real data 95 per cent of the time in simulations, then
it is ruled out at 95 per cent confidence.
For a T2 model, we can rule out a length scale of
0.3 (in units of the comoving distance to the last-scattering surface)
for the small dimensions at 95 per cent confidence.  For
T1 models, the limits are somewhat weaker: there
does not appear to be a 95 per cent confidence limit, but 
a length scale of 0.2 is ruled out at 90 per cent confidence.
As we might expect in light of the comments above, these
limits are weaker than those set by other more focused techniques
such as the $S$ statistic.  But they are quite general, and could also
be used to constrain any other family of models which possess
directionality.

\section{Utility}
Since the next generation of satellite experiments will produce
extremely large
(mega-pixel) data sets, it is important to devise statistical
techniques that can be rapidly and efficiently calculated.
This is an especially important concern in cases where
the statistic has to be calibrated with Monte Carlo simulations.
The statistic ${\cal D}$ described in this paper can be evaluated
for an $N_{\rm p}$-pixel map with $O(N_{\rm p})$ 
calculations,
which
means that extensive Monte Carlo simulations will be possible
even with mega-pixel maps.

The slowest step in the calculation of ${\cal D}$ is estimating
the gradient map.  Suppose the map has $N_{\rm p}$ pixels divided
into $N_{\rm b}$ blocks, so that there are $M\equiv N_{\rm p}/N_{\rm b}$
pixels per block.  Then each gradient estimate
involves fitting a linear function to $M$ data
points, at a cost of $O(M^2)$ operations.  The total operations count
is therefore $O(M^2N_{\rm b})$, or $O(MN_{\rm p})$, with a prefactor of order
unity.  The appropriate value of $M$ in future experiments is unlikely
to be much greater than the value $M=64$ used above: indeed, it
may prove to be smaller, since future maps will have lower pixel
noise than \COBE.  It should therefore be possible to estimate $\cal D$ 
with an operation count of order $100N_{\rm p}$.
Even for data sets as large as those expected from \MAP\ and \Planck, it will be
possible to calculate the
statistic, and calibrate it with simulations, quite rapidly.

This statistic may also prove useful for sky maps at other wavelengths,
such as x-ray and infrared maps.  It can also be generalised
to apply to maps of point sources by pixelizing the map in pixels
large enough to contain many sources each.  Thus it could be applied to
galaxy or star catalogues which cover a sufficiently large solid angle.

The technique described in this paper can provide a valuable test for
Galactic or zodiacal contamination in sky maps.  Suppose
we calculate
$\cal D$ for a map from which such signals are
supposed to have been removed.  If $\cal D$ proves
to be large, and if the maximum or minimum value of
$f$ occurs in a direction close to the Galactic or ecliptic pole,
then we would have good reason for suspecting residual contamination.

The $\cal D$ statistic can also naturally be applied to polarisation
maps.  In fact, the application to a polarisation map is even simpler
than the technique described in this paper, since one can simply skip
the initial gradient estimation and use the polarisation measurements
themselves in place of the gradient estimates.  (Note that the
function $f$ depends quadratically on $\vecg_p$, so the fact that a
polarisation vector is defined only up to a sign does not matter.)
Since the cosmic polarisation signals for \MAP\ and \Planck\ are expected
to be rather weak, and the galactic signals fairly uncertain on large
scales, the $\cal D$ statistic may prove useful in checking future
polarisation maps for residual galactic contamination.

\section{Conclusions}
When the \MAP\ and \Planck\ satellites return their high-fidelity maps of the
sky, they will be subjected to the full battery of available statistical tests.
Specific patterns will be searched for, in order to constrain particular
models, but there will also be general tests for patterns on the sky or
general sorts of non-Gaussianity which might be present.  Among those
basic properties to search for, directionality is quite fundamental, and
could be connected with various models for the structure of the Universe
on the largest scales.  We have discussed a general method for finding
directionality, the $\cal D$ statistic, which requires no particular
assumptions about the specific form of the directional pattern.  This
is a rather obvious statistic, being quadratic in the gradient of the
temperature field, and with the only complication being the derivation
of the appropriate weights to use when the map is not all-sky.

For polarisation maps, the application of the $\cal D$ statistic is even more
straightforward, because the gradient does not have to be taken.  And
since the statistic is quite general, it could obviously be applied to
maps at other wavelengths.  For large-angle catalogues of discrete objects
(e.g.~x-ray clusters or radio galaxies) this statistic could also be
utilised, after first extracting a smooth field from the catalogue.  As
well as providing constraints on tera-parsec scale exotica, there is a much
more mundane use for the $\cal D$ statistic, namely testing for
residual foreground or other systematic effects.  For large maps our
statistic will be particularly easy to apply, since it is straightforward
and fast to calculate, a great advantage given the sizes of data set
which are about to deluge us.

\section*{ACKNOWLEDGMENTS}
DS was supported by the Natural Sciences and Engineering
Research Council of Canada.

\end{document}